\documentclass[twocolumn,showpacs,preprintnumbers,amsmath,amssymb,aps,prd,superscriptaddress]{revtex4-1}
\makeatletter

\newcommand{\Rmnum}[1]{\expandafter\@slowromancap\romannumeral #1@}
\makeatother
\usepackage[colorlinks,linkcolor=blue,anchorcolor=blue,citecolor=blue,urlcolor=blue,breaklinks=true]{hyperref}

\usepackage{epsfig}
\usepackage{natbib}
\usepackage{epstopdf}
\usepackage{mathrsfs}
\usepackage{slashed}
\usepackage{amsmath}
\usepackage{verbatim}
\usepackage{graphicx}
\usepackage{amssymb}
\usepackage{psfrag}
\usepackage{array}
\usepackage{lipsum}
\usepackage{float}
\usepackage[normalem]{ulem}
\usepackage[dvipsnames]{xcolor}

\begin{document}

\title{Deep learning for jet modification in the presence of the quark gluon plasma background}
\author{Ran Li}
\affiliation{Institute of Frontier and Interdisciplinary Science, Shandong University, Qingdao, 266237, China}
\author{Yi-Lun Du}
\email{yilun.du@iat.cn}
\affiliation{Shandong Institute of Advanced Technology, Jinan 250100, China}
\author{Shanshan Cao}
\email{shanshan.cao@sdu.edu.cn}
\affiliation{Institute of Frontier and Interdisciplinary Science, Shandong University, Qingdao, 266237, China}
\begin{abstract}
Jet interactions with the color-deconfined QCD medium in relativistic heavy-ion collisions are conventionally assessed by measuring the modification of the distributions of jet observables with respect to their baselines in proton-proton collisions. Deep learning methods enable per-jet evaluation of these modifications, enhancing the use of jets as precision probes of the nuclear medium. In this work, we predict the jet-by-jet fractional energy loss $\chi$ for jets evolving through a quark-gluon plasma (QGP) medium using a Linear Boltzmann Transport (LBT) model. To approximate realistic experimental conditions, we embed medium-modified jets in a thermal background and apply Constituent Subtraction for background removal. Two network architectures are studied: convolutional neural networks (CNNs) using jet images, and dynamic graph convolutional neural networks (DGCNNs) using particle clouds. We find that CNNs achieve accurate predictions for background-free jets but degrade in the presence of the QGP background and remain below the background-free baseline even after background subtraction. In contrast, DGCNNs applied to background-subtracted particle clouds maintain high accuracy across the entire $\chi$ range, demonstrating the advantage of point-cloud–based graph neural networks that exploit full jet structure under realistic conditions.

\end{abstract}

\maketitle


\section{Introduction} 

Quark-Gluon Plasma (QGP) is a novel state of matter formed at extremely high temperature and/or density~\cite{Gyulassy:2004zy, Jacobs:2004qv, Busza:2018rrf, Elfner:2022iae}. In this state, partons are no longer confined within hadrons and can move freely over distances significantly larger than typical hadronic sizes. This state can be created in relativistic heavy-ion collisions, providing a controlled environment for studying strongly interacting matter under extreme conditions. Hard processes occurring in the early stage of relativistic heavy-ion collisions produce high-transverse-momentum partons, which undergo showering and fragmentation to form nearly collimated streams of final-state particles, known as jets~\cite{PhysRevLett.68.1480, Qin:2015srf, Cao:2020wlm}.
When these energetic partons traverse the QGP medium, they interact with the medium's constituents, leading to energy loss or even complete absorption into the medium. This phenomenon is referred to as jet quenching, which has been extensively studied experimentally to probe the properties of the QGP.
For instance, the observed suppression of jet spectra~\cite{Abelev:2013kqa,Adam:2015ewa,ATLAS:2014ipv,Aaboud:2017eww,ATLAS:2018gwx,Acharya:2019jyg} in nucleus-nucleus collisions reflects energy loss induced by quenching effects. Furthermore, comprehensive investigations on modifications of jet substructures~\cite{Apolinario:2024equ, Casalderrey-Solana:2016jvj, Tachibana:2017syd, KunnawalkamElayavalli:2017hxo, Milhano:2017nzm, Chen:2020tbl, Park:2018acg, Luo:2018pto, Casalderrey-Solana:2019ubu, Chang:2019sae, Tachibana:2020mtb, Yang:2023dwc, Xing:2024yrb} in nuclear collisions provide critical insight into the influence of quenching on the internal dynamics of jets.

Despite rapid progress, jet quenching studies continue to face challenges. The steeply falling jet spectra with respect to transverse momentum ($p_\mathrm{T}$) make jet modification measurements highly susceptible to selection bias~\cite{Brewer:2018dfs,Du:2020pmp,du2021jet,Takacs:2021bpv,Brewer:2021hmh,Andres:2024hdd}: when comparing jet samples from proton-proton and nucleus-nucleus collisions within the same $p_{\mathrm{T}}$ interval, jets that have lost less energy are more likely to be selected.
To minimize the selection bias, an ideal approach is to estimate the energy loss on a jet-by-jet basis. However, the quenching process experienced by jets in the medium is highly complex: the path length traversed by each jet varies, and local medium properties such as temperature and flow velocity further influence the energy loss. Moreover, the large and fluctuating background inherent in heavy-ion environments poses additional difficulties for jet studies.

Among different approaches proposed to address the challenge of selection bias, machine learning (ML) offers a unique capability to characterize the energy loss process at the single-jet level.
ML technologies have brought significant progress to jet physics. They have been successfully applied in various topics, including quark/gluon jet classification~\cite{Chien:2018dfn,du2021classification}, quenched jet identification~\cite{apolinario2021deep,Liu:2022hzd,lai2021information,romao2023jet,qureshi2024model}, jet production vertex localization~\cite{yang2023deep}, and background subtraction~\cite{Haake:2018hqn,ALICE:2023waz,Li:2024fzn,Stewart:2024mkx,Qureshi:2025ylv}. In parallel, diverse jet representations have been explored, including image-based~\cite{baldi2016jet,de2016jet,komiske2017deep,kasieczka2017deep}, sequence-based~\cite{Guest:2016iqz,dreyer2018lund}, tree-based~\cite{louppe2019qcd,cheng2018recursive}, and point-cloud representations~\cite{Komiske:2018cqr,Dolan:2020qkr,PhysRevD.101.056019,Guo:2020vvt,Ju:2020tbo,Gong:2022lye}.

In previous studies~\cite{Du:2020pmp,du2021jet}, per-jet energy loss was predicted using machine learning methods based on a hybrid strong/weak coupling model~\cite{casalderrey2015erratum}, without including effects of the QGP background particles. A crucial step toward applying such methods to real experimental data, however, is to assess their robustness in realistic heavy-ion environments where jets are embedded in a large fluctuating background. In this work, we explicitly introduce thermal background particles on top of the simulated jets to emulate experimental conditions, and employ constituent subtraction to mitigate their impact. In addition, we employ the Linear Boltzmann Transport (LBT) model~\cite{cao2016linearized,Luo:2023nsi} to simulate jet evolution in the medium, thereby not only extending the predictive framework to more realistic scenarios but also testing its applicability across different jet quenching models. Using this setup, we demonstrate that ML techniques, specifically convolutional neural networks (CNNs) and dynamic graph convolutional neural networks (DGCNNs)~\cite{wang2019dynamic}, can effectively predict the jet-by-jet energy loss ratio, offering a promising path to mitigate the selection bias in jet quenching studies.

The remainder of this paper is structured as follows. In Sec.~\ref{sec: sim models}, we describe the simulation frameworks: PYTHIA for proton-proton collisions, LBT for jet–medium interactions, and a thermal toy model for the QGP background, as well as a jet matching procedure used to define the energy loss ratio. Section~\ref{sec:const} details the Constituent Subtraction method for background removal. Section~\ref{sec: ML} introduces the ML methodologies, where CNNs process jet images and DGCNNs operate on particle-cloud representations, dynamically learning the graph structure of jet constituents. In Sec.~\ref{sec:performance}, we present the prediction performance of these ML methods: CNNs achieve accurate predictions for clean jets, while under background conditions, DGCNNs consistently outperform CNNs across the entire energy loss ratio range. Finally, Sec.~\ref{sec: Conclusions} summarizes our findings and discusses future directions.

\section{Simulation models}
\label{sec: sim models}

We use PYTHIA~\cite{Sjostrand:2006za,Sjostrand:2014zea} to generate jet partons produced in proton-proton ($p+p$) collisions. They also serve as the initial-state partons in nucleus-nucleus (A+A) collisions prior to scatterings through the QGP. The Linear Boltzmann Transport (LBT) model~\cite{cao2016linearized,Luo:2023nsi} is employed to simulate jet interactions with the QGP. In order to study impacts of the QGP background on jet reconstruction, we use a toy thermal model~\cite{JetToyHI} to mimic thermal (soft) particle production from the QGP. Below, we briefly summarize the key features of these models and their parameter setups in the present study.

PYTHIA is a well established event generator of high-energy particle collisions. We use it to simulate the production of energetic (hard) partons in nucleon-nucleon collisions, and the subsequent splittings of these partons, known as parton showers, at their high-virtuality stage. By default, splittings of a parton cease when it reaches a scale of 0.5~GeV. In this work, we generate 140k $p+p$ collision events at a center-of-mass energy per nucleon pair of $\sqrt{s_\mathrm{NN}}=5.02$~TeV using PYTHIA~8 with the Monash 2013 tune~\cite{Skands:2014pea}. Triggers of momentum exchange of hard collisions are uniformly sampled within [300, 1000]~GeV. We apply the FastJet~\textcolor{black}{3.3.1} toolkit~\cite{Cacciari:2011ma} on the final state partons from the PYTHIA shower to reconstruct them into jets for $p+p$ collisions. Meanwhile, these partons can further interact with the QGP before they are reconstructed into quenched jets for A+A collisions.

The LBT model simulates elastic and inelastic scatterings between low-virtuality jet partons and thermal partons inside the QGP medium based on the Boltzmann equation. The elastic scattering rates of jet partons are calculated by convoluting the $2\rightarrow 2$ matrix elements at the leading order with thermal distributions of the medium partons at a given temperature, with all partonic scattering channels taken into account~\cite{He:2015pra}; and the inelastic scattering rates are related to the number of medium-induced gluons per unit time, which are taken from the higher-twist energy loss calculations~\cite{Guo:2000nz,Wang:2001ifa,Zhang:2003wk,Majumder:2009ge}. According to these scattering rates, the LBT model not only updates the kinematic information of jet partons as they plough through the medium, but also sample ``recoil" partons scattered out of the medium by jets, leaving energy holes (named as ``negative" partons) inside the medium. Recoil partons are treated equally to jet partons originating from the PYTHIA shower, and can undergo further scatterings with the medium. Contributions from negative partons to jet observables need to be subtracted from that from regular jet partons and recoil partons. These recoil and negative partons constitute jet-induced medium excitation (or medium response) in the LBT model, which is essential to guarantee the energy-momentum conservation of the jet+medium system and improve our quantitative understanding of jet observables in relativistic heavy-ion collisions. The medium information, including its local temperature and flow velocity, is provided by the (3+1)-dimensional viscous hydrodynamic model CLVisc~\cite{Pang:2018zzo,Wu:2021fjf}. 

In this work, we simulate the environment created in Pb+Pb collisions at $\sqrt{s_\mathrm{NN}}=5.02$~TeV within the 0-10\% centrality bin. Positions of hard nucleon-nucleon collisions are sampled using the Monte-Carlo Glauber model. Each jet parton produced by PYTHIA is assumed to stream freely until both its formation time, defined as the sum of the splitting times of its ancestors in the PYTHIA shower~\cite{Zhang:2022ctd}, and the thermalization time of the QGP (0.6~fm). Then, jet partons interact with the hydrodynamic medium until they exit the QGP, i.e., when the local temperature of the medium falls below 165~MeV. The strong coupling constant is set as $\alpha_\mathrm{s}=0.3$ in LBT. The output partons from LBT are then clustered into jets for A+A collisions. Due to the lack of a reliable hadronization scheme for jet partons in A+A collisions, we use partons from PYTHIA and LBT, instead of hadrons, to reconstruct jets for both $p+p$ and A+A collisions. This will be improved in our upcoming efforts.

To study the performance of ML models in realistic experimental environments, we also employ a thermal toy model~\cite{JetToyHI} to sample soft particles emitted by the QGP background. For each A+A collision event, the thermal model is tuned to generate $\pi^\pm$ particles with momenta following a Boltzmann distribution that gives a total multiplicity of 1538 and an average transverse momentum of $\langle p_\mathrm{T} \rangle = 0.5696$~GeV within the rapidity range of $|y| < 1$, corresponding to the soft hadron spectra created in 0-10\% Pb+Pb collisions at $\sqrt{s_\mathrm{NN}} = 5.02$~TeV~\cite{acharya2020production}. 

We will embed jet partons from LBT into these background particles and reconstruct jets with and without background subtraction. This will allow us to investigate how the QGP background affects the ML models' accuracy in predicting the jet energy loss. 
For convenience, we refer to jets reconstructed without the QGP background as ``LBT-only jets" or "background-free LBT jets", those reconstructed with the QGP background as ``LBT jets within background", and those reconstructed after background subtraction as ``LBT jets with background subtracted". For all samples, jets are reconstructed using the anti-$k_\mathrm{T}$ algorithm~\cite{Cacciari:2008gp} with a radius parameter of $R = 0.4$. For analysis, we select jets with transverse momenta $p_\mathrm{T} >100$~GeV and pseudorapidity within $|\eta| < 1-R$. 

The primary objective of this work is to determine, on a jet-by-jet basis, the amount of energy lost by jets as they propagate through the QGP medium. The energy loss ratio is quantified by a variable~\cite{Du:2020pmp,du2021jet}
\begin{equation}
\label{eq:chi-definition}
\chi \equiv \frac{p_\mathrm{T}^f}{p_\mathrm{T}^i} ,
\end{equation}
where $p_\mathrm{T}^f$ denotes the transverse momentum of a jet after traversing the medium, simulated using the LBT model, and $p_\mathrm{T}^i$ corresponds to the transverse momentum of the \emph{same} jet in the absence of medium effects, simulated using PYTHIA. For LBT jets embedded in the QGP background, the $p_\mathrm{T}^f$ used in the definition of $\chi$ includes contributions only from partons generated by LBT, excluding contributions from the background particles.

Establishing this definition requires a dedicated matching between a quenched LBT jet ($j_\mathrm{LBT}$) and its unquenched PYTHIA counterpart ($j_\mathrm{PYTHIA}$). For each $j_\mathrm{LBT}$, whether embedded in the background, background-subtracted, or not, we search for its counterpart $j_\mathrm{PYTHIA}$ from its corresponding PYTHIA event. The distance between $j_\mathrm{LBT}$ and $j_\mathrm{PYTHIA}$ is required to satisfy $\Delta R = \sqrt{\Delta\eta^2 + \Delta\phi^2} < 0.4$. If multiple PYTHIA jets meet this criterion, the hardest one is chosen as the match. Conversely, if no PYTHIA jet satisfies the condition, the $j_\mathrm{LBT}$ candidate is excluded from both training and analysis. Starting with the 140k PYTHIA events, we obtain a final dataset of about 170k medium-modified jets for this study. 


\section{Constituent Subtraction Method}
\label{sec:const}

To mitigate possible degradation of the prediction performance of ML models on the jet energy loss ratio due to the contamination from the QGP background, a background subtraction procedure may be essential to recover or improve the performance. We employ the Constituent Subtraction (CS) method~\cite{Berta:2019hnj,JetToyHI}, which operates directly at the particle level. Instead of correcting jets after clustering, this method iteratively modifies or removes the momenta of individual particles before reconstruction. Such an approach is particularly well suited for heavy-ion environments, where the abundance of soft particles can distort clustering and bias jet observables~\cite{cacciari2008catchment}.

The procedure begins with an estimate of the average background momentum density $\rho$. This is obtained as the median of $p_{\mathrm{T}, i}/A_i$ over clusters reconstructed with the $k_\mathrm{T}$ algorithm~\cite{Catani:1993hr,Ellis:1993tq}, excluding the few hardest clusters (two in this work) to avoid bias from real jets.
Here, $p_{\mathrm{T}, i}$ is the transverse momentum of the $i$-th cluster and $A_i$ its geometric area, determined via the ghost-particle method. To mimic the underlying event, a dense set of ghost particles is distributed uniformly across the rapidity-azimuthal-angle ($y$-$\phi$) plane. Each ghost is assigned a transverse momentum $p_\mathrm{T}^g = \rho A^g$, where $A^g=0.001$ denotes its effective area, so that the ensemble of ghosts reproduces the overall background contribution.

\begin{figure}[tbp!]
\centering
\includegraphics[width=0.49\textwidth]{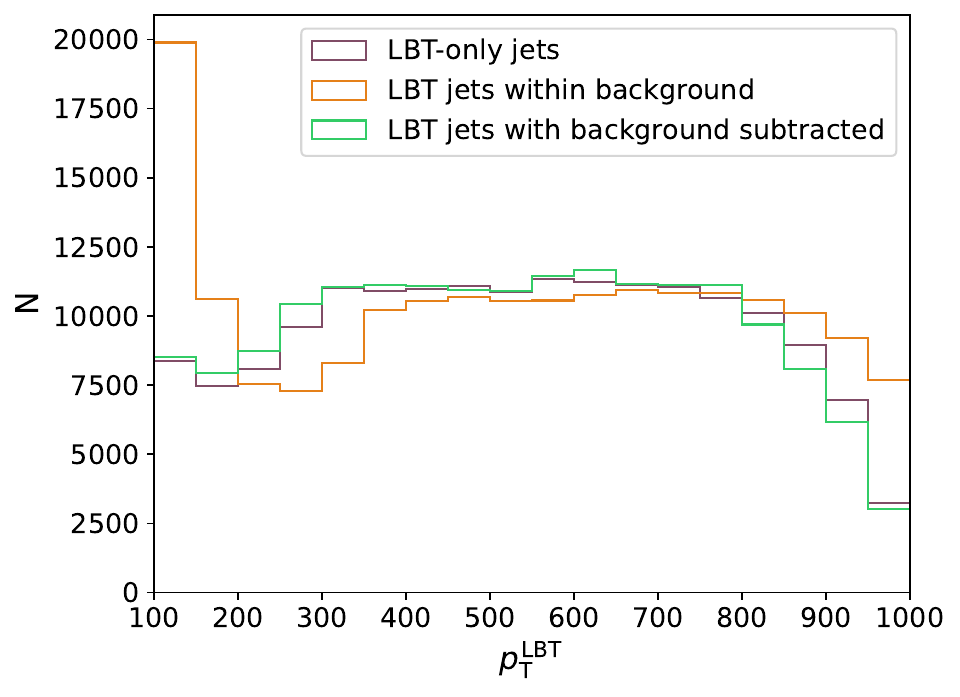}\\
\caption{(Color online) Transverse momentum distributions of the LBT-only jets, the LBT jets within background, and the LBT jets with background subtracted.}
\label{fig: CS performance pT} 
\end{figure}

Pairs are then formed between each real particle $i$ and ghost $k$, with their distance defined as
\begin{equation}
\Delta R_{i,k} = p_{\mathrm{T},i}^{\alpha} \sqrt{(y_i - y_k^g)^2 + (\phi_i - \phi_k^g)^2},
\end{equation}
where $\alpha$ is a tunable parameter, set to $1$ in this work. This definition incorporates the $p_\mathrm{T}$ weighting, favoring the removal of softer particles. Distances are ordered from the smallest to the largest, and momentum subtraction is carried out pair by pair:
$$
\begin{array}{l}
p_{\mathrm{T}, i} \to p_{\mathrm{T}, i} - p_{\mathrm{T}, k}^{g}, \quad p_{\mathrm{T}, k}^{g} \to 0, \quad \text{if} \quad p_{\mathrm{T}, i} \geq p_{\mathrm{T}, k}^{g};\\
p_{\mathrm{T}, k}^{g} \to p_{\mathrm{T}, k}^{g} - p_{\mathrm{T}, i},   \quad  p_{\mathrm{T}, i} \to 0,\quad \text{otherwise.} 
\end{array}
$$ 
The procedure continues until the distance between pairs exceeds $\Delta R^{\mathrm{max}}$, which sets the subtraction range. We use $\Delta R^{\mathrm{max}} = \infty$ in our analysis. After subtraction, jets are reconstructed with the anti-$k_\mathrm{T}$ algorithm using only particles that retain nonzero momentum.

By locally correcting each particle according to the estimated background density, Constituent Subtraction succeeds in removing a large fraction of the soft background while preserving the essential jet kinematics and substructure. This makes it particularly effective for heavy-ion jet analyses.

\begin{figure}[tbp!]
\centering
\includegraphics[width=0.49\textwidth]{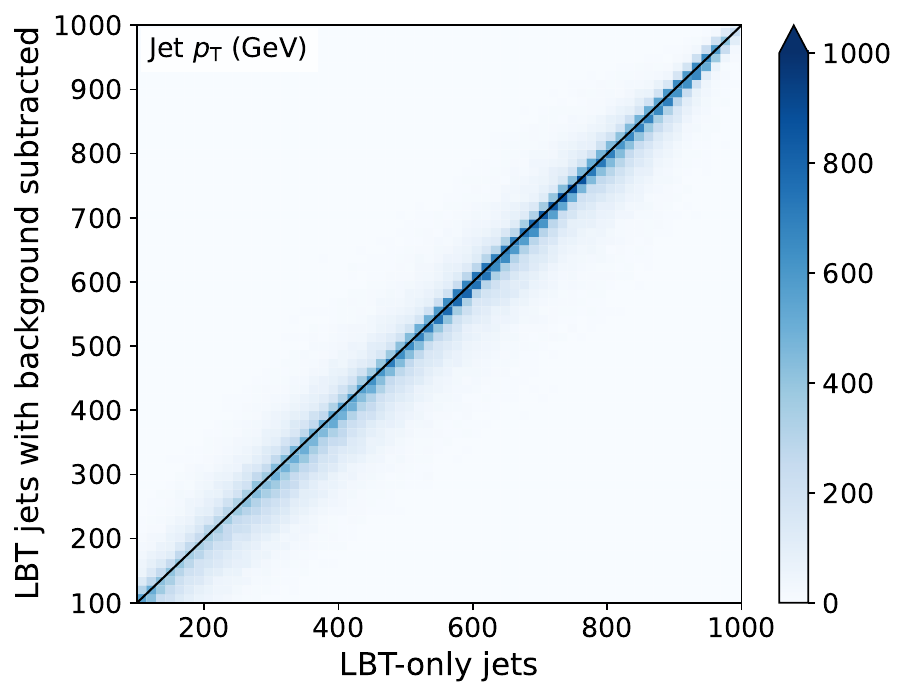}\\
\caption{(Color online) Joint $p_\mathrm{T}$ distribution of the LBT-only jets and the background-subtracted LBT jets.}
\label{fig: CS pT compare} 
\end{figure}

\begin{figure}[tbp!]
\centering
\includegraphics[width=0.49\textwidth]{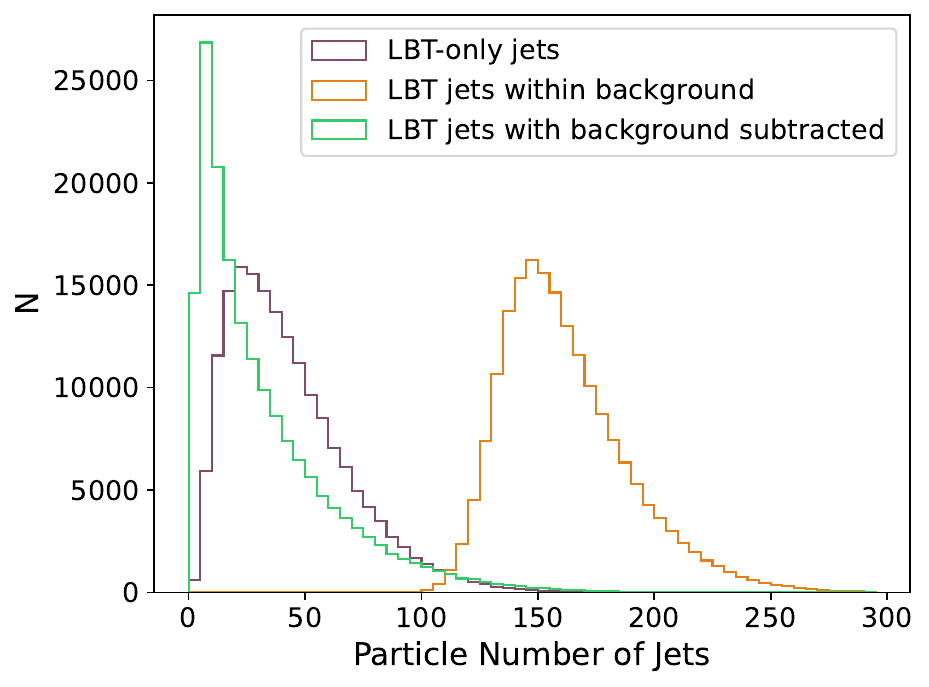}\\
\caption{(Color online) Particle number distributions of jets, compared between the LBT-only jets, the LBT jets within background, and the LBT jets with background subtracted.}
\label{fig: CS performance N} 
\end{figure}

To scrutinize the performance of the background subtraction, in Fig.~\ref{fig: CS performance pT}, we compare the transverse momentum distributions between three jet categories: the LBT-only (background-free) jets, jets within background, and jets with background subtracted via Constituent Subtraction, demonstrating that the subtraction procedure effectively recovers the $p_\mathrm{T}$ distributions of the LBT-only jets. On the jet-by-jet level, Fig.~\ref{fig: CS pT compare} presents the joint $p_\mathrm{T}$ distribution of the LBT-only jets and the background-subtracted LBT jets. The matching procedure introduced at the end of Sec.~\ref{sec: sim models} is used to pair the background-subtracted LBT jet sample with the LBT-only jet sample. The close alignment between the two samples here further confirms the accuracy of the Constituent Subtraction method in recovering the jet $p_\mathrm{T}$ in the presence of the QGP background. 
In Fig.~\ref{fig: CS performance N}, we compare the particle number distributions of jets between the same three jet categories. One can clearly observe the clustering of background particles into jets when jet partons are embedded into the QGP background. This can be effectively corrected using the Constituent Subtraction method, though  a slight over-subtraction in the particle multiplicity is hinted.

\section{Machine learning methods}
\label{sec: ML}
In this work we use two machine learning architectures in order to extract the fractional energy loss $\chi$ on a jet-by-jet basis. These are the convolutional neural network (CNN) and the dynamic graph convolutional neural network (DGCNN).

\subsection{Convolutional Neural Network}
\label{subsec:CNN}

\begin{figure*}[tbp!]
\centering
\includegraphics[width=0.999\textwidth]{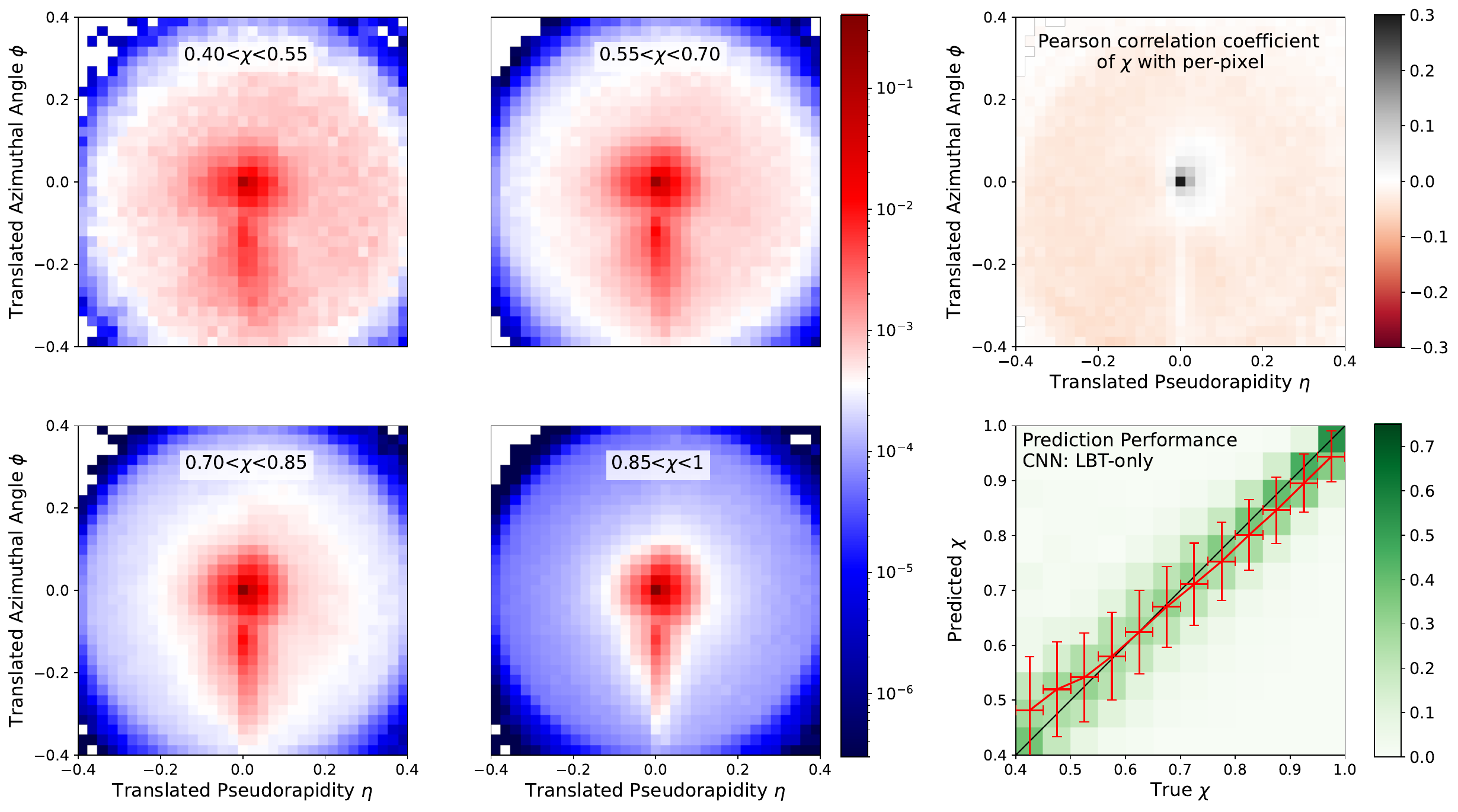}
\caption{(Color online) The left four panels show the average images from self-normalized energy distributions inside the LBT jets reconstructed without introducing the QGP background particles, grouped into four energy loss ratio ($\chi$) bins. The top right panel presents the Pearson correlation coefficient between $\chi$ and value at each pixel of unnormalized jet image. The bottom right panel presents the CNN prediction performance on $\chi$ obtained using unnormalized jet images.}
\label{fig: Jet_image_no_bkg} 
\end{figure*}

We adopt the same CNN architecture as in Ref.~\cite{Du:2020pmp} (see also Refs.~\cite{pang2018equation,du2020identifying} for technical details). The network consists of three convolutional layers (16, 16, and 32 filters of sizes $8\times8$, $7\times7$, and $6\times6$) followed by a fully connected layer with 128 neurons, and a single output neuron predicting the energy loss ratio $\chi$. All convolutional and dense layers use “He normal” initialization~\cite{he2015delving}, L2 regularization~\cite{ng2004feature}, batch normalization~\cite{ioffe2015batch}, PReLU activation~\cite{he2015delving}, and Dropout (0.2 after convolutions, 0.5 after the dense layer). Average pooling ($2\times 2$) follows the first and third convolutional layers.

The network is trained as a regression model using the Log-Cosh loss and optimized with AdaMax~\cite{kingma2014adam} at a learning rate of $10^{-4}$. Training uses batches of 1024 jet images, reshuffled each epoch, for up to 400 epochs with checkpointing at the best validation loss.

The input jet image $J(\eta,\phi)$ is a $33\times33$ matrix of the total $p_\mathrm{T}$ deposited per pixel, covering translated pseudorapidity $|\eta|\leq 0.4$, and translated azimuthal angle $|\phi|\leq 0.4$ for a fixed jet radius $R=0.4$. Preprocessing follows Refs.~\cite{de2016jet,komiske2017deep}. We first translate the jet images so the hardest groomed subjet is at $(0,0)$. Then we rotate it so the second-hardest subjet is at $-\pi/2$ (or align the first principal component if absent). Finally we apply a parity flip so the right half has larger total intensity.

\subsection{Dynamic Graph Convolutional Neural Network}
\label{subsec:DGCNN}

To improve prediction accuracy, we adopt the ParticleNet architecture~\cite{PhysRevD.101.056019}, which is based on the dynamic graph convolutional neural network (DGCNN)~\cite{wang2019dynamic}. This model operates directly on jet constituents represented as an unordered point cloud, stacking multiple EdgeConv modules and dynamically updating neighborhood graphs between layers using the $k$-nearest neighbors (KNN) algorithm~\cite{fix1952discriminatory,cover1967nearest}.

In ParticleNet, the EdgeConv module processes two types of input information for each particle: its “coordinates” in a chosen feature space and its associated high-dimensional feature vector. For each particle, the $k$ nearest neighbors are identified based on Euclidean distances in the coordinate space ($k=16$ in this work). The features of each particle and its neighbors are combined to form edge features that encode both local geometry and particle-level information. These edge features are transformed by edge functions and subsequently aggregated to update the representation of each particle.

The architecture consists of three EdgeConv modules, each parameterized by a multi-layer perceptron (MLP) with channel dimensions (64, 64, 64), (128, 128, 128), and (256, 256, 256), respectively. The first module computes distances in the $(\Delta\eta,\Delta\phi)$ plane relative to the jet axis, using the particle four-momentum as input features. The subsequent modules take the learned features from the previous stage as both new coordinates and features. After the EdgeConv layers, a global average pooling operation aggregates the features into a representation of size 256, which is passed through a fully connected layer with 256 neurons and ReLU activation~\cite{glorot2011deep}. To mitigate overfitting, a Dropout layer with rate 0.1 is applied. The final output layer has two units with Softmax normalization, yielding probabilities that sum to one, with one component interpreted as the predicted energy-loss ratio.

The network is trained using the AdamW optimizer~\cite{loshchilov2019fixing} with a learning rate of $10^{-4}$. Training is performed with batches of 1024 jet point clouds, reshuffled at each epoch, for up to 10 epochs, with model checkpointing based on the best validation loss.

\begin{figure*}[tbp!]
\centering
\includegraphics[width=0.999\textwidth]{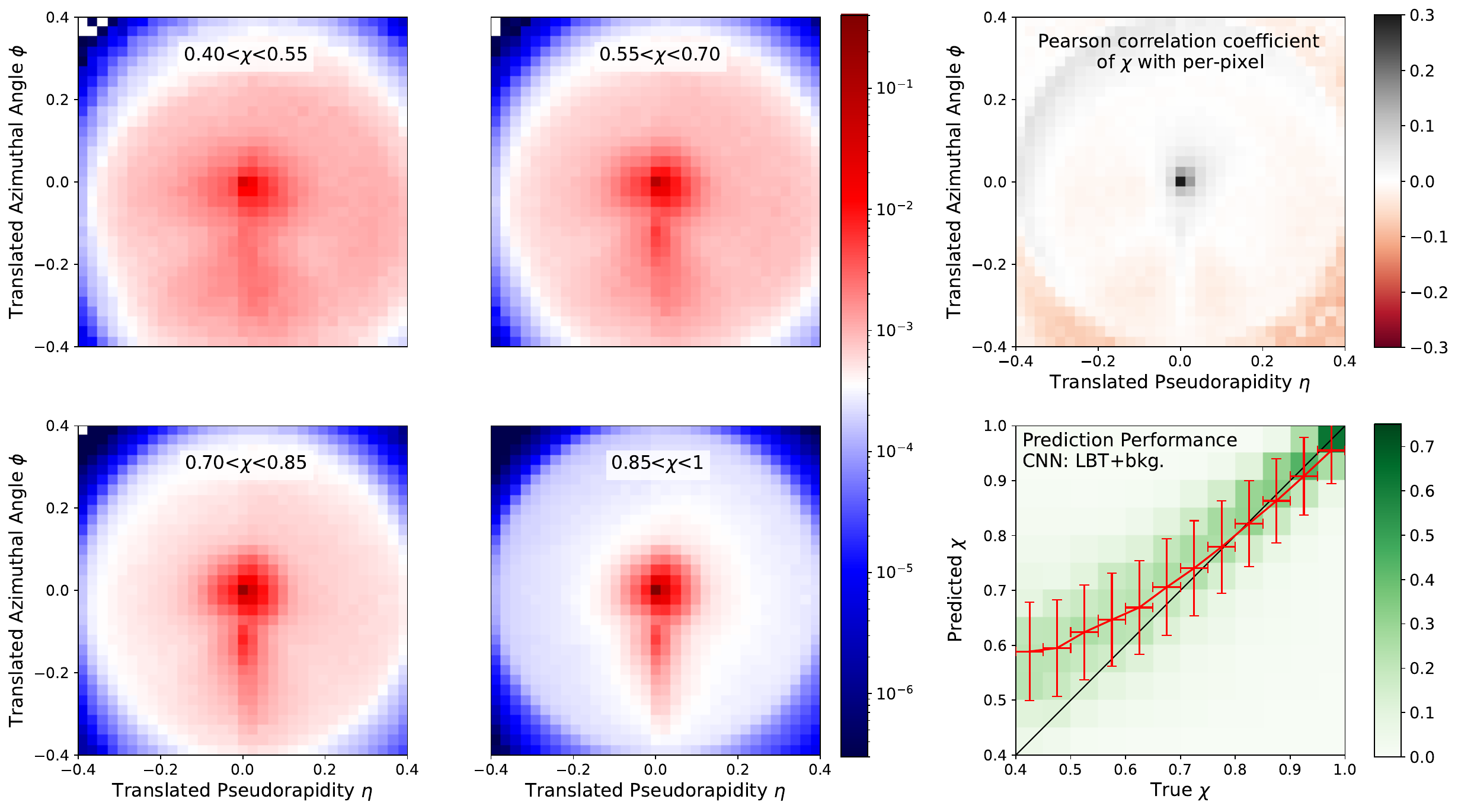}
\caption{(Color online) Same as Fig.~\ref{fig: Jet_image_no_bkg}, except that jets are reconstructed together with the QGP background particles.
}
\label{fig: Jet_image_with_bkg} 
\end{figure*}

The architectural advantage of DGCNN lies in its ability to capture fine-grained substructure patterns that are inherently lost in image-based representations. While CNN's pixelization process blurs these precise particle-level relationships by compressing continuous kinematic information into discrete bins, DGCNN's EdgeConv operations preserve geometric fidelity through direct computation of feature differences between neighboring particles. Furthermore, the dynamic graph update mechanism allows the network to hierarchically integrate local constituent information, enabling it to learn both detailed particle-level correlations and larger-scale jet substructure patterns with higher fidelity.

Since the DGCNN involves dynamic $k$-nearest-neighbor searches and multiple EdgeConv operations, it requires more computational time than the CNN used in this work. On a single NVIDIA T4 GPU, the training times of our tasks are about 8 hours for the CNN and 33 hours for the DGCNN. The per-jet inference times are about 0.05 ms for the CNN and 0.5 ms for the DGCNN.

\section{Prediction performance}
\label{sec:performance}

We first study the performance of the CNN on predicting jet energy loss with images of energy distribution inside jets as input. To facilitate feature extraction by the CNN, we identify leading and subleading sub-jets inside each jet using the Soft Drop algorithm~\cite{Larkoski:2014wba} with parameters $z_\mathrm{cut}=0.1$ and $\beta=0$, and translate the leading sub-jet to the center of the $\eta$-$\phi$ plane, rotate the subleading one to the $-\pi/2$ direction, and reflect the image if necessary to ensure that its right half owns higher intensity than its left half. Shown in the left four panels of Fig.~\ref{fig: Jet_image_no_bkg} are images of jets within different ranges of the energy loss ratio $\chi$, with smaller $\chi$ representing stronger jet energy loss. Here, jets are reconstructed using the output particles from the LBT simulation, without contribution from the background particles of the QGP, i.e., the ``LBT-only" or "background-free" setup defined in Sec.~\ref{sec: sim models}. To demonstrate the features of quenched jets visually, 
we average self-normalized jet images in each bin of $\chi$ per panel, where the normalization prevents high-$p_\mathrm{T}$ jets from dominating the results.
From these average jet images, one can observe the emergence of soft particles at large angles relative to the hard core of a jet, represented by its leading sub-jet, as the jet scatters through the QGP. The jet energy appears more sparsely distributed as it loses more energy, or $\chi$ is smaller. This could be a distinct correlation between jet energy loss and jet image grasped by the CNN. 
To quantify this correlation, in the top right panel of Fig.~\ref{fig: Jet_image_no_bkg} we present the Pearson correlation coefficient between $\chi$ and the value at each pixel of unnormalized jet image. The Pearson coefficient between samples $X$ and $Y$ with $n$ population is defined as 
\begin{equation}
r_{xy} =\frac{\sum ^n _{i=1}(x_i - \bar{x})(y_i - \bar{y})}{\sqrt{\sum ^n _{i=1}(x_i - \bar{x})^2} \sqrt{\sum ^n _{i=1}(y_i - \bar{y})^2}}.
\end{equation}
This coefficient quantifies the linear correlation between two quantities, ranging from $-1$ (perfect anti-correlation) to $+1$ (perfect correlation), with 0 indicating no linear correlation. Large positive values of the coefficient are observed near the jet core, reflecting a positive correlation between $\chi$ and the remaining energy of the hard core of a jet. In contrast, negative values are observed at large angles, reflecting stronger energy transport from the jet core towards large angles for more quenched jets (with smaller $\chi$). 
This further illustrates the potential of using a CNN to infer the amount of jet energy loss from their images.

\begin{figure*}[tbp!]
\centering
\includegraphics[width=0.999\textwidth]{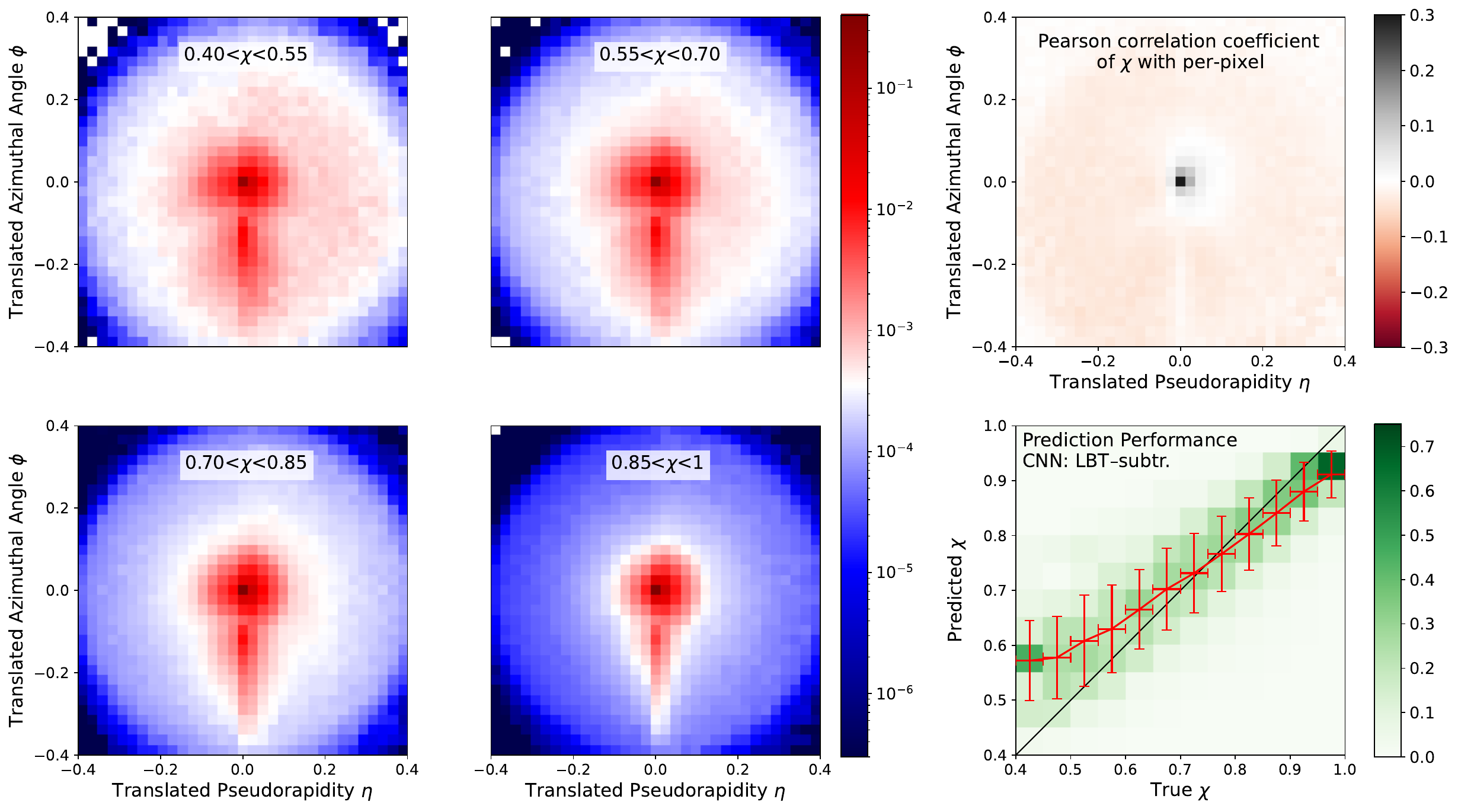}
\caption{(Color online) Same as Figs.~\ref{fig: Jet_image_no_bkg} and~\ref{fig: Jet_image_with_bkg}, except that the QGP background particles are introduced and then subtracted before jet reconstruction.
}
\label{fig: Jet_image_remove_bkg} 
\end{figure*}

Using the unnormalized background-free jet images as input, we train the CNN model as described in Sec.~\ref{sec: ML} to predict the jet energy loss ratio $\chi$. The bottom right panel of Fig.~\ref{fig: Jet_image_no_bkg} shows the prediction performance, with the horizontal axis representing the true value of $\chi$ and the vertical axis representing the CNN prediction. The green shading shows the conditional probability distribution of the predicted $\chi$ given a true value, normalized per true-$\chi$ bin. The red line with error bars shows the mean and standard deviation of predictions in each true-$\chi$ bin. The mean prediction closely follows the diagonal, indicating that the CNN effectively learns the mapping from jet image to the energy loss ratio $\chi$ for the LBT jets in the absence of the QGP background.

To assess the prediction performance under more realistic experimental conditions, we embed the LBT output particles in a soft-particle background before jet reconstruction. The background particles are generated using a thermal model, as described in Sec.~\ref{sec: sim models}, and this setup is referred to as ``LBT jets within background" (``LBT+bkg."). The corresponding average jet images are shown in the left four panels of Fig.~\ref{fig: Jet_image_with_bkg}. Compared to the background-free case shown above, a large number of background particles are clustered into jets, as evident in Fig.~\ref{fig: CS performance N}. This contamination becomes more significant when jets lose more energy. This is because for more quenched jets, more recoil particles are generated at large angles, which can be hardly distinguished from background particles in jet reconstruction. Therefore, compared to the left four panels of Fig.~\ref{fig: Jet_image_no_bkg}, we see more sparsely distributed jet energy in  Fig.~\ref{fig: Jet_image_with_bkg} when jet energy loss is strong. For significantly quenched jets, e.g., in the range of $0.4 < \chi < 0.7$, the variation of jet images with respect to $\chi$ can be barely observed. The weaker correlation between $\chi$ and jet image is also reflected in the Pearson coefficients in the top right panel, which get closer to zero inside the jet cone ($R<0.4$) compared to those in Fig.~\ref{fig: Jet_image_no_bkg}. Interestingly, negative Pearson coefficients emerge outside the $R=0.4$ cone here, indicating considerable numbers of background particles outside the cone are clustered into jets when jet-medium interaction is strong. When these jet images are used as input to the CNN, the prediction performance, shown in the bottom right panel of Fig.~\ref{fig:  Jet_image_with_bkg}, exhibits clear deviations from the diagonal for $\chi \lesssim 0.7$. Thus, the degradation of the image-level features caused by background contamination directly translates into the reduced predictive power of the CNN. These results highlight the necessity of background subtraction before feeding jet images to machine learning models.

\begin{figure}[tbp!]
\centering
\includegraphics[width=0.49\textwidth]{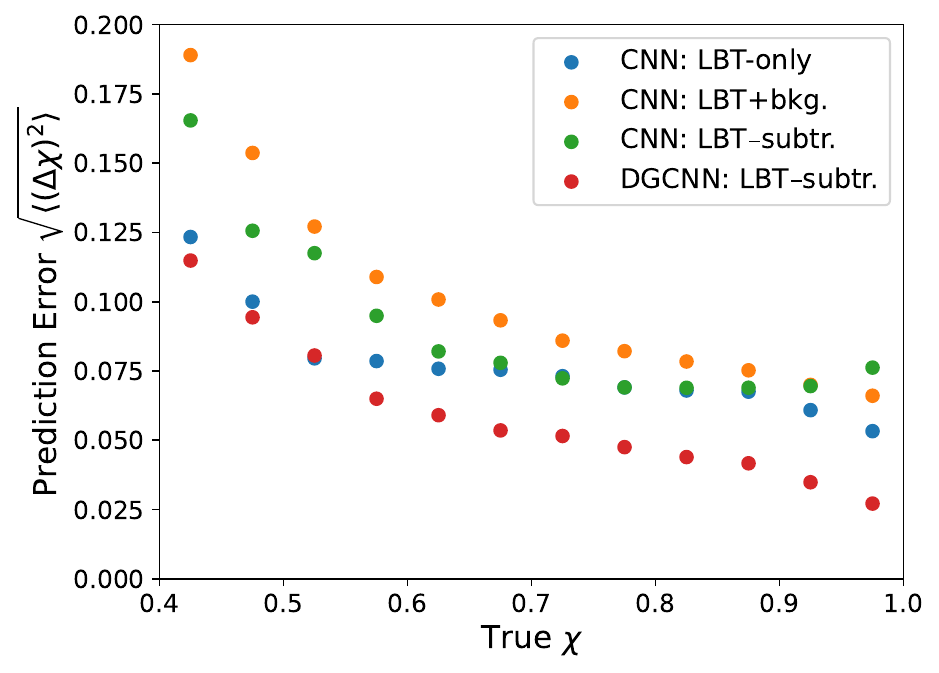}
\caption{(Color online) Prediction errors $\sqrt{\langle (\Delta \chi)^2 \rangle}$ as functions of the true value of $\chi$, compared between the CNNs trained using three jet samples (LBT-only, LBT+bkg., LBT-subtr.) and the DGCNN trained using the LBT-subtr. jets.}
\label{fig: compare performance} 
\end{figure}

Now we apply background subtraction to the LBT jets embedded in the background using the Constituent Subtraction method as described in Sec.~\ref{sec:const}. The corresponding results are labelled as ``LBT jets with background subtracted" (``LBT-subtr."). As shown in Fig.~\ref{fig: Jet_image_remove_bkg}, the average jet images (left four panels) largely recover the $\chi$-dependent features of jet quenching seen in Fig.~\ref{fig: Jet_image_no_bkg}, and negative values of Pearson coefficients re-emerge clearly inside the jet cone (top right panel). The mean CNN predictions for these images (bottom right panel) remain close to the diagonal, though slight deviations persist across the entire $\chi$ range. 

For a quantitative comparison between different model setups, we show in Fig.~\ref{fig: compare performance} their prediction errors $\sqrt{\langle (\Delta \chi)^2 \rangle}$ as functions of the true value of $\chi$. The error is defined for each true-$\chi$ bin as the root-mean-square deviation between the true and predicted values of $\chi$: $\sqrt{\langle (\Delta \chi)^2 \rangle}=\sqrt{\sum_{i\in \mathrm{bin}}(\chi^\mathrm{true}_i-\chi^\mathrm{pred}_i)^2/N_\mathrm{bin}}$, where $N_\mathrm{bin}$ is the number of jets in the bin. The prediction accuracy based on the LBT jets with background subtracted improves for $\chi\lesssim0.90$ but degrades for $\chi>0.95$ compared to the case based on that within background. Nevertheless, the overall performance remains below that of the background-free (LBT-only) case, likely due to an over-subtraction of soft particles by the Constituent Subtraction method, as illustrated in Fig.~\ref{fig: CS performance N} and the comparison between jet images in Figs.~\ref{fig: Jet_image_no_bkg} and~\ref{fig: Jet_image_remove_bkg}. These results indicate that although background subtraction partially restores the predictive power of CNNs, significant limitations persist, motivating the need for alternative representations beyond jet images.

\begin{figure}[tbp!]
\centering
\includegraphics[width=0.49\textwidth]{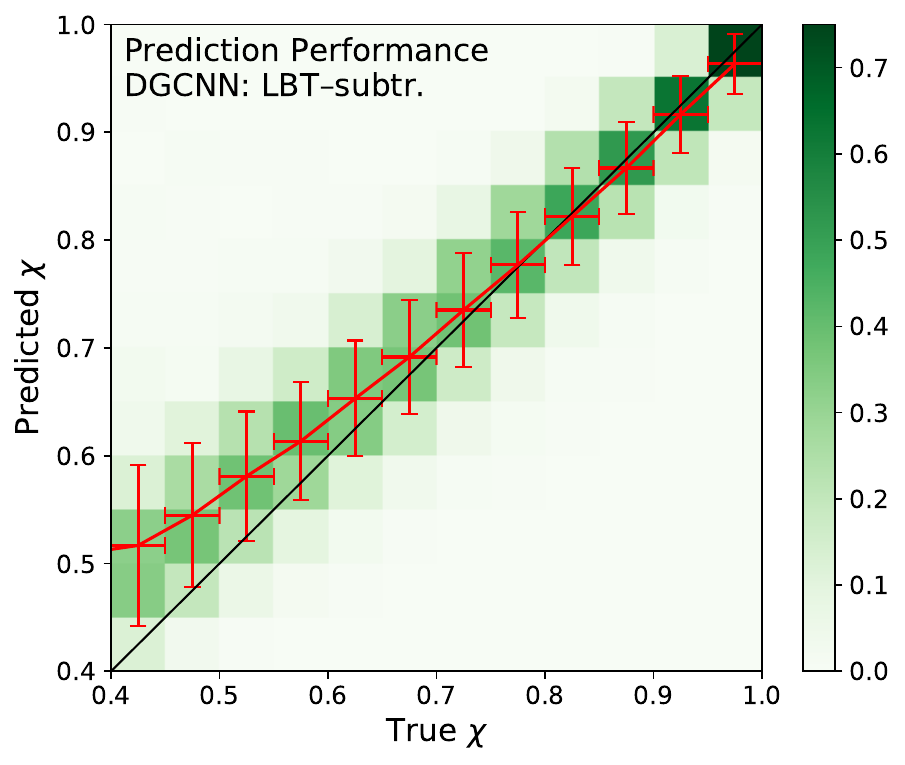}
\caption{(Color online) The prediction performance of the DGCNN with the LBT jets with background subtracted as input. The green shading represents the probability of the predicted $\chi$ along the $y$-axis given a true value of $\chi$. Each column is normalized here. The red line and error bars quantify the mean and standard deviation of the predicted $\chi$, respectively,  given the true $\chi$.}
\label{fig: dgcnn performance} 
\end{figure}

To further improve the prediction performance, we replace the CNN with a DGCNN. As discussed in Sec.~\ref{subsec:DGCNN}, for each background-subtracted LBT jet, we generate a point-cloud representation and use it as  input to the DGCNN. Figure~\ref{fig: dgcnn performance} shows the corresponding prediction results. Within each true-$\chi$ bin, the predicted values are more tightly clustered around the mean, and the mean line lies closer to the diagonal compared to the previous CNN results, indicating a more accurate per-bin prediction. The prediction error as a function of the true $\chi$, quantified in Fig.~\ref{fig: compare performance}, further demonstrates the advantage of the DGCNN: the root-mean-square deviation $\sqrt{\langle (\Delta \chi)^2 \rangle}$ is smaller across the full $\chi$ range compared to the CNN results, even relative to the background-free case. We note that although the performance of the CNN could potentially be improved by further refining the background subtraction procedure, our main conclusion regarding the superior performance of the DGCNN remains unchanged. In particular, the ``LBT-subtr.” setup of the DGCNN outperforms the ``LBT-only” setup of the CNN, with the latter corresponding to an idealized scenario of perfect background subtraction that is not achievable in realistic experiments. This demonstrates that the advantage of the DGCNN is robust even in the presence of realistic fluctuations introduced by background contamination and subtraction.

It is important to understand why the point-cloud–based DGCNN architecture achieves a higher prediction accuracy than the image-based CNN. A previous study \cite{Du:2020pmp} showed that a fully connected network trained on a set of jet-shape variables, fragmentation functions, and Soft-Drop substructure observables can reach a performance comparable to CNNs trained on jet images. This indicates that CNNs are indeed capable of learning features closely related to these physically motivated observables, even though such information is partly blurred by the pixelization of jet images.

In contrast, the point-cloud representation used by DGCNN preserves the full per-particle kinematic information without discretization, enabling the network to access more detailed correlations among jet constituents. Through the EdgeConv operations, the DGCNN directly processes local geometric relationships between neighboring particles, which allows it to capture fine-grained structures such as total particle multiplicity and its angular distribution that are strongly correlated with medium-induced jet modifications. While it is inherently difficult to determine which specific jet features are learned by a neural network, the improved performance of the DGCNN suggests that models operating on minimally processed particle-level data can more faithfully encode the jet–medium interaction patterns that govern the energy-loss fraction $\chi$.

\section{Conclusions} 
\label{sec: Conclusions}

We implemented two neural-network architectures, convolutional neural networks (CNNs) and dynamic graph convolutional neural networks (DGCNNs), to predict the jet-by-jet energy loss ratio $\chi$ in LBT simulations, aiming to reduce the selection bias in jet quenching studies. Extending earlier work~\cite{Du:2020pmp} based on hybrid strong/weak coupling models~\cite{casalderrey2015erratum} for background-free jets, we demonstrated that these machine-learning approaches remain effective for jets quenched in the LBT model, even in the presence of the QGP background. Each jet can be represented in two ways: as an energy-deposition image for the CNN, and as a particle cloud for the DGCNN. The jet images intuitively reveal that stronger quenching corresponds to enhanced soft-particle activity at large angles, while the particle-cloud representation retains full constituent-level information.

For the background-free LBT jets, the CNN effectively predicts the true $\chi$ values. To approximate realistic experimental conditions, we embedded jets in a soft-particle background, and then applied background subtraction before training. CNN performance degrades in the presence of background and remains below the background-free baseline even after the background removal with the Constituent Subtraction method. In contrast, the DGCNN applied to background-subtracted particle clouds achieves consistently higher accuracy across the entire $\chi$ range ($0.4$-$1.0$), demonstrating that point-cloud-based methods, by leveraging the full jet structure, retain superior predictive power for jet energy loss even after background contamination and removal.

To reliably apply our framework for per-jet energy loss predictions under realistic experimental conditions, future studies must address five key directions. First, it is necessary to incorporate hadronization into our jet modeling so that the trained neural networks can be applied more reliably to real data. While hadronization is not expected to significantly affect the jet images used in the CNN, it does influence the per-particle information that the DGCNN relies on. Second, incorporating more realistic background particles into the simulations is essential for faithfully reproducing experimental environments. This includes sampling soft hadrons from hydrodynamic simulations of the QGP and embedding non-jet particles from hard nucleon–nucleon scatterings to model the underlying event. In addition, future studies should consider pile-up and detector effects to more accurately reflect realistic analyses. Third, developing background subtraction techniques that better preserve jet substructures will be crucial to mitigate distortions from the soft QGP particles. Fourth, it is important to validate, and if necessary, enhance the generalizability of machine learning approaches across different jet quenching models. Although we expect the ML models trained using the LBT data should retain certain predictive power when applied to jets generated from other quenching models, as long as those models contain proper descriptions of jet-medium interactions and capture the key correlation between jet energy loss and jet structure, a more reliable generalizability would require training the networks on a combined dataset from multiple jet models. Last but not least, Fig.~\ref{fig: compare performance} shows that the prediction errors generally increase for jets that experience larger energy loss, regardless of which ML model is used and whether background contamination is present or subtracted. This trend reflects the fact that strongly quenched jets tend to exhibit more complex internal structures in heavy-ion collisions. As confirmed in an earlier study~\cite{Du:2020pmp}, the prediction errors grow for jets with large groomed jet radius $R_g$, large Soft Drop multiplicity $n_{SD}$, and (groomed) jet mass around 40 GeV. These observations motivate the exploration of more advanced architectures or alternative ML techniques capable of capturing the subtler features of medium-modified jet structure, thereby further improving both the precision and robustness of per-jet energy-loss predictions.

Finally, we comment on the definition of the jet energy-loss fraction used throughout this work. Our study employs the inclusive definition
$\chi = p_T^{\mathrm{jet,med}}/p_T^{\mathrm{jet,vac}}$,
based on matched jets with $\Delta R < 0.4$, which naturally incorporates possible modifications of the jet axis and constituent composition during in-medium evolution. This definition provides a direct and physically transparent measure of the total energy loss in theoretical jet-quenching studies. Nonetheless, jets in realistic heavy-ion environments can experience significant background contamination and medium response, motivating the exploration of more differential and potentially more robust quenching measures. Examples include groomed-jet $p_T$ fractions, leading-subjet $p_T$ fractions, and jet core energy fractions, which emphasize the hardest branches or the jet core and therefore reduce sensitivity to soft background activity. Although such alternatives were not investigated in the present work, systematically examining groomed-based or subjet-based definitions of the energy-loss fraction thus constitutes an interesting and promising direction for future research.

\section*{Acknowledgements}
We are grateful for helpful discussions with Xiao-Zhi Bai. This work is supported in part by the National Natural Science Foundation of China (NSFC) under Grant Nos. 12175122, 2021-867, 12321005 (R.L., S.C.), and in part by the Taishan Scholars Program under Grant No. tsqnz20221162, and Shandong Excellent Young Scientists Fund Program (Overseas) under Grant No. 2023HWYQ-106 (Y.D.).


\bibliographystyle{apsrev4-1}
\bibliography{duyl}



\end{document}